\documentclass[twocolumn,aps,showpacs,preprintnumbers,amsmath,amssymb,superscriptaddress]{revtex4}

\usepackage{graphicx}
\usepackage{color}

\usepackage{array}

\begin{document}

\newcommand{\bra}[1]{\left\langle {#1} \right|}
\newcommand{\ket}[1]{\left|  #1 \right\rangle}
\newcommand{\bracket}[3]{\langle {#1} | {#2} | {#3} \rangle}
\newcommand{\braket}[2]{\langle {#1} | {#2} \rangle}
\newcommand{\aver}[1]{\langle {#1} \rangle}
\newcommand{\abs}[1]{\left| {#1} \right|}
\newcommand{\oper}[1]{\hat{#1}}
\newcommand{\coper}[1]{\hat{#1}^\dag}
\newcommand{\raw}[0]{\rightarrow}

\newcommand{\xunit}{\ensuremath{\mathbf{\hat x}}}
\newcommand{\yunit}{\ensuremath{\mathbf{\hat y}}}
\newcommand{\zunit}{\ensuremath{\mathbf{\hat z}}}

\newcommand{\mirrortrans}{\ensuremath{q}}


\title{Simplified measurement of the Bell parameter within quantum mechanics}

\author{Haruka Tanji}
\affiliation{Department of Physics, MIT-Harvard Center for
Ultracold Atoms, and Research Laboratory of Electronics,
Massachusetts Institute of Technology, Cambridge, Massachusetts
02139, USA}
\affiliation{Department of Physics, Harvard University,
Cambridge, Massachusetts 02138, USA}

\author{Jonathan Simon}
\affiliation{Department of Physics, MIT-Harvard Center for
Ultracold Atoms, and Research Laboratory of Electronics,
Massachusetts Institute of Technology, Cambridge, Massachusetts
02139, USA}
\affiliation{Department of Physics, Harvard University,
Cambridge, Massachusetts 02138, USA}

\author{Saikat Ghosh}
\affiliation{Department of Physics, MIT-Harvard Center for
Ultracold Atoms, and Research Laboratory of Electronics,
Massachusetts Institute of Technology, Cambridge, Massachusetts
02139, USA}

\author{Vladan Vuleti\'{c}}
\affiliation{Department of Physics, MIT-Harvard Center for
Ultracold Atoms, and Research Laboratory of Electronics,
Massachusetts Institute of Technology, Cambridge, Massachusetts
02139, USA}

\date{\today}

\begin{abstract}
We point out that, if one accepts the validity of quantum mechanics, the
Bell parameter for the polarization state of two photons can be
measured in a simpler way than by the standard procedure
[Clauser, Horne, Shimony, and Holt, Phys.\ Rev.\ Lett.\ {\bf 23}, 880
(1969)]. The proposed method requires only two measurements with
parallel linear-polarizer settings for Alice and Bob at $0^\circ$ and
$45^\circ$, and yields a significantly smaller statistical error for a large Bell parameter.
\end{abstract}

\maketitle

Measurements of the Bell parameter $S$ for a pair of two-level
systems, and verification of Bell's inequality $|S| \leq 2$,
were originally designed to experimentally distinguish
local hidden-variable theories from quantum mechanics
\cite{Bell66,Clauser69}. Many experiments performed since have
shown that it is possible to violate Bell's inequality, ruling
out local hidden-variable theories, and confirming quantum
mechanics
\cite{Freedman72,Aspect81,Aspect82,Aspect82a,Weihs98,Rowe01}.
Nevertheless, numerous Bell measurements, with photons
\cite{Marcikic04}, ions \cite{Rowe01}, or combinations of photons
and particles \cite{Blinov04a}, continue to be performed. With the
validity of quantum mechanics already established (except for the
\emph{simultaneous} \cite{Cabello07,Brunner07} closing of the
locality \cite{Weihs98} and detection \cite{Rowe01} loopholes),
most such measurements serve primarily to confirm the
non-classical character of two-particle states. In particular,
in quantum cryptography, measurements of the Bell parameter are
used to test if the secret key has been compromised
\cite{Ekert91}.

\begin{figure}
\includegraphics[width=6.5 cm, angle=-90]{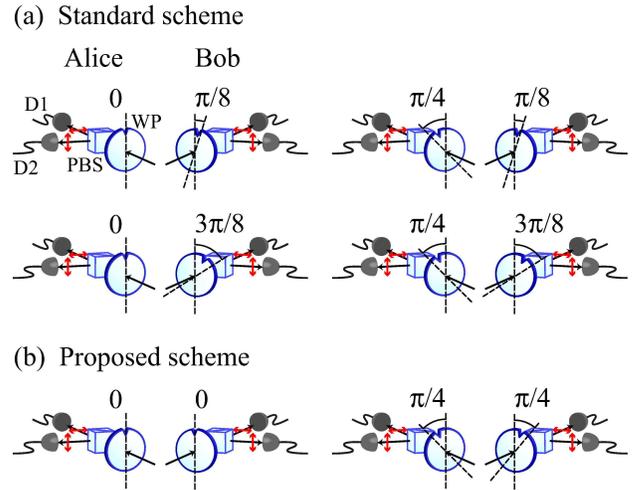}
\caption{(Color online) 
(a) Bell parameter measurement for photon polarizations 
according to Clauser, Horne, Shimony, and Holt (CHSH), and (b)
the proposed alternative scheme. Alice and Bob each receive a photon,
and measure its linear polarization.  
Each detection path consists of a half-wave plate (WP), a polarizing beam splitter cube (PBS), and 
two detectors (D1 and D2) for the two orthogonal polarizations.
The labeled angles are the analysis angles, which are twice the wave plate settings.
(a) In the CHSH scheme, Alice
and Bob perform measurements with four combinations of angles
($\alpha_1=0$ and $\alpha_2=\pi/4$ for Alice,
$\beta_1=\pi/8$ and $\beta_2=3\pi/8$ for Bob),
to determine four correlation coefficients $E(\alpha_i,\beta_j)$.
They then calculate the Bell parameter as $S_\pm=\pm
\left[E(0, \pi/8)-E(0,3\pi/8)\right]+
E(\pi/4,\pi/8)+E(\pi/4,3\pi/8) $.
(b) Assuming the validity of quantum mechanics, Alice and Bob can
simply use two identical angles, $\alpha^\prime_1=\beta^\prime_1=0$ and
$\alpha^\prime_2=\beta^\prime_2=\pi/4$, and determine the Bell
parameter as $S^\prime_\pm=\sqrt{2} \left[ \pm E(0,0) +
E(\pi/4,\pi/4) \right]$.} \label{setup}
\end{figure}

The standard procedure to verify Bell's inequality for photon
polarization states, introduced by Clauser, Horne, Shimony and
Holt (CHSH) \cite{Clauser69}, is as follows.
The recipients of the two photons, Alice and Bob, measure coincidences for four
combinations of linear-polarizer angles:  Alice measures at
angles $\alpha_1=0$ and $\alpha_2=\pi/4$, and Bob at
$\beta_1=\pi/8$ and $\beta_2=3\pi/8$ (see Fig.~\ref{setup}).
The Bell parameter is then calculated as
\begin{equation}
\label{CHSH} S_\pm = \pm \left[
E(\alpha_1,\beta_1)-E(\alpha_1,\beta_2) \right] +
E(\alpha_2,\beta_1)+E(\alpha_2,\beta_2),
\end{equation}
where
\begin{equation}
E(\alpha_i,\beta_j)=p_+(\alpha_i,\beta_j)-p_-(\alpha_i,\beta_j)
\label{CorrCoeff}
\end{equation}
is the correlation coefficient of the measurement
$\{\alpha_i,\beta_j\}$. Here $p_+(\alpha_i,\beta_j)$ denotes the
fraction of events where the polarization measurements by Alice at
angle $\alpha_i$ and by Bob at $\beta_j$ are positively correlated
(both photons pass through their respective polarizers, or both are rejected) 
and $p_-(\alpha_i,\beta_j)$ denotes
the fraction of events where the photons are anticorrelated (one
passes the polarizer, and the other is rejected). If the
photons are perfectly correlated, then $E(\alpha_i,\beta_j)=+1$;
for perfectly anticorrelated photons, we have
$E(\alpha_i,\beta_j)=-1$. States with $|S|>2$ violate the notion
of local realism \cite{Bell66}, 
while the maximum magnitude of the Bell parameter $S$ that quantum mechanics allows
is $|S|=2\sqrt{2}$. Experiments with photons
\cite{Aspect81,Aspect82,Aspect82a}, and later with ions
\cite{Rowe01,Blinov04a}, have shown that systems yielding $|S|>2$
can indeed be prepared, with measured values as large as
$|S|=2\sqrt{2}-0.003(19)$ \cite{Wong06}.

The CHSH procedure, using Eq.~(\ref{CHSH}), allows one to experimentally
rule out local hidden-variable theories \cite{Bell66,Clauser69}.
If one merely attempts to characterize
\cite{Altepeter05,Khan04} a two-qubit state within quantum
mechanics by the Bell parameter, however, the question arises whether it is
\emph{necessary} that Alice and Bob perform four measurements, and
along different polarization axes. Assuming the validity of quantum mechanics, 
we show that it is in fact sufficient for Alice and Bob
to measure in the same polarization basis, for two different bases
(e.g., $\alpha^\prime_1=\beta^\prime_1=0$ and
$\alpha^\prime_2=\beta^\prime_2=\pi/4$), to determine the Bell parameter
$S^\prime_\pm= S_\pm$ more simply as
\begin{equation}
\label{simpleS} S^\prime_\pm = \sqrt{2} \left[ \pm E(\alpha^\prime_1,\beta^\prime_1) +
E(\alpha^\prime_2,\beta^\prime_2) \right].
\end{equation}
As discussed below, such a prescription arises naturally from the
transformation properties of Bell states under rotations. 
The proposed method discriminates directly
between all four Bell states via the signs in Eq.~(\ref{simpleS}),
and requires only two combinations of polarizer settings whereas the
CHSH scheme requires four.
Finally, our method provides significantly smaller statistical
errors for states that violate Bell's inequality. 

We should note that, while Eq.~(\ref{simpleS}) represents a simpler measurement of the Bell
parameter within quantum mechanics, $S^\prime_\pm$ in this particular form 
does not rule out local hidden-variable theories.
In other words, $S^\prime_\pm=S_\pm$ holds within quantum
mechanics, but not generally in local hidden-variable theories.
There is currently an interesting dispute 
\cite{Christian07,Grangier07} as to whether
Bell's theorem in a broader sense is valid, or whether
non-commuting local realistic models \cite{Christian07} can invalidate 
the conclusions drawn from standard Bell tests
\cite{Bell66,Clauser69,Freedman72,Aspect81,Aspect82,Aspect82a,Rowe01,Weihs98}
about the non-local or non-realistic character of quantum
mechanics. 
If the latter is the case, or if one merely wishes to characterize two-photon states, 
it is preferable to use the quantum-mechanically equivalent, 
simplified form of the Bell parameter, Eq.~(\ref{simpleS}), 
rather than the CHSH prescription, Eq.~(\ref{CHSH}).

The states that yield the maximum violation of Bell's inequality
$|S_\pm| =2\sqrt{2} \nleq 2$, are the Bell states
$\ket{\Phi_{\pm}^{HV}}$, $\ket{\Psi_{\pm}^{HV}}$, obtained by
applying the Bell-state creation operators
\begin{eqnarray} \label{BellStateOperators}
\left( \oper{\Phi}_{\pm}^{HV} \right)^\dag  &=& \frac{\coper{h}_A
\coper{h}_B \pm \coper{v}_A \coper{v}_B}{\sqrt{2}} \nonumber \\
\left( \oper{\Psi}_{\pm}^{HV} \right)^\dag  &=& \frac{\coper{h}_A
\coper{v}_B \pm \coper{v}_A \coper{h}_B}{\sqrt{2}},
\end{eqnarray}
to the photon vacuum $\ket{0}$,
\begin{eqnarray} \label{BellStatesHV}
\ket{\Phi_{\pm}^{HV}} &\equiv& \left( \oper{\Phi}_{\pm}^{HV} \right)^\dag \ket{0} = \frac{ \ket{hh} \pm \ket{vv}}{\sqrt{2}} \nonumber  \\
\ket{\Psi_{\pm}^{HV}} &\equiv& \left( \oper{\Psi}_{\pm}^{HV}
\right)^\dag \ket{0} = \frac{ \ket{hv} \pm \ket{vh}}{\sqrt{2}}.
\end{eqnarray}
Here $\ket{ab} \equiv \ket{a}_A \ket{b}_B$ denotes the
polarizations $a$, $b$ of the photons received by Alice (A) and Bob (B),
respectively. We have defined $\ket{h}_A = \coper{h}_A \ket{0}$,
with $\coper{h}_A$ being the creation operator for one
horizontally polarized photon in Alice's mode, with similar
definitions for $\ket{v}_A, \ket{h}_B$ and $\ket{v}_B$. The $HV$
superscript indicates that these Bell states are defined in the
horizontal/vertical ($HV$) basis.

We now consider a linear polarization basis $ST$ (for both Alice
and Bob) at $45^\circ$ relative to the $HV$ basis, defined by the
photon creation operators
\begin{eqnarray} \label{45DegreeBasis}
\coper{s}_\mathcal{K} &=& \frac{1}{\sqrt{2}} \left(
\coper{h}_\mathcal{K} + \coper{v}_\mathcal{K}
\right) \nonumber \\
\coper{t}_\mathcal{K} &=& \frac{1}{\sqrt{2}} \left(
-\coper{h}_\mathcal{K} + \coper{v}_\mathcal{K} \right),
\end{eqnarray}
where $\mathcal{K}=A,B$. 
It is then easy to see the correspondence of the Bell states between the two bases:
\begin{eqnarray} \label{BellStatesHVST}
\ket{\Phi_+^{ST}} &\equiv& \frac{ \ket{ss} + \ket{tt}}{\sqrt{2}} = 
\frac{ \ket{hh} + \ket{vv}}{\sqrt{2}} = \ket{\Phi_+^{HV}}\nonumber \\
\ket{\Phi_-^{ST}} &\equiv& \frac{ \ket{ss} - \ket{tt}}{\sqrt{2}} =
\frac{ \ket{hv} + \ket{vh}}{\sqrt{2}} = \ket{\Psi_+^{HV}}\nonumber \\
\ket{\Psi_+^{ST}} &\equiv& \frac{ \ket{st} + \ket{ts}}{\sqrt{2}} =
\frac{ -\ket{hh} + \ket{vv}}{\sqrt{2}} = -\ket{\Phi_-^{HV}}\nonumber \\
\ket{\Psi_-^{ST}} &\equiv& \frac{ \ket{st} - \ket{ts}}{\sqrt{2}} =
\frac{ \ket{hv} - \ket{vh}}{\sqrt{2}} = \ket{\Psi_-^{HV}}.
\end{eqnarray}
The above relations show that, e.g., the Bell state
$\ket{\Phi_+^{HV}}$ has maximum positive correlation in the $HV$
basis (Alice and Bob always measure the same polarization), as
well as in the $ST$ basis, while the state $\ket{\Phi_-^{HV}}$
displays maximum positive correlation in the $HV$ basis, and
maximum anticorrelation in the $ST$ basis. In contrast, a mixed
state with equal probability of $\ket{hh}$ and $\ket{vv}$ pairs
shows no correlations in the $ST$ basis. Therefore one may suspect
that the Bell states can be identified by the combination of
correlation measurements in the $HV$ and $ST$ bases, and that it
may not be necessary for Alice and Bob to measure at nonzero
relative polarizer angles.

To demonstrate the equivalence of Eqs.~(\ref{CHSH}) and (\ref{simpleS})
for measuring the Bell parameter, we show that both expressions
represent the expectation value of the same Bell operator
\cite{Braunstein92}. For an arbitrary (pure or mixed) photon-pair
input state as described by a density matrix $\rho$, the
correlation coefficient $E(\alpha_i,\beta_j)$ can be written as
the expectation value $E(\alpha,\beta)=\mathrm{Tr}\left( \rho
\oper{\mathcal{E}}_{\alpha \beta} \right)$ of the operator
\begin{eqnarray}\label{EOperator}
\oper{\mathcal{E}}_{\alpha \beta} =
\coper{\alpha}\oper{\alpha}\coper{\beta}\oper{\beta} +
\coper{\alpha}_{\perp}\oper{\alpha}_{\perp}\coper{\beta}_{\perp}\oper{\beta}_{\perp}
- \coper{\alpha}_{\perp}\oper{\alpha}_{\perp}
\coper{\beta}\oper{\beta}
-\coper{\alpha}\oper{\alpha}\coper{\beta}_{\perp}\oper{\beta}_{\perp}
\nonumber \\  = \left[\left( \coper{h}_A \oper{h}_A - \coper{v}_A
\oper{v}_A \right) \cos 2\alpha + \left( \coper{h}_A \oper{v}_A +
\coper{v}_A \oper{h}_A \right) \sin 2\alpha  \right] \nonumber \\
\times \left[\left(\coper{h}_B \oper{h}_B - \coper{v}_B \oper{v}_B
\right) \cos 2\beta + \left( \coper{h}_B \oper{v}_B + \coper{v}_B
\oper{h}_B \right) \sin 2\beta \right].
\end{eqnarray}
Here $\oper{\alpha} = \oper{h}_A \cos \alpha + \oper{v}_A \sin
\alpha$ and $\oper{\beta} = \oper{h}_B \cos \beta + \oper{v}_B
\sin \beta$ are the annihilation operators for a photon polarized
along angles $\alpha$ in Alice's and $\beta$ in Bob's mode,
respectively, while $\alpha_{\perp}=\alpha+\pi/2$ and
$\beta_{\perp}=\beta+\pi/2$ denote the corresponding orthogonal polarizations.
Eq.~(\ref{EOperator}) follows directly from the definition of the
correlation coefficient [Eq.~(\ref{CorrCoeff})], since
$\coper{a}_A \oper{a}_A \coper{b}_B \oper{b}_B$ measures the
fraction of photon pairs with polarizations $a$, $b$,
observed by Alice and Bob, respectively. The Bell parameter is the
expectation value $S_\pm=\mathrm{Tr}\left( \rho \oper{S}_\pm \right)$ of the operator
\begin{equation}\label{BellOperator}
\oper{S}_{\pm} = \pm \left( \oper{\mathcal{E}}_{0,\frac{\pi}{8}} -
\oper{\mathcal{E}}_{0,\frac{3\pi}{8}} \right) +
\oper{\mathcal{E}}_{\frac{\pi}{4},\frac{\pi}{8}}
+\oper{\mathcal{E}}_{\frac{\pi}{4},\frac{3\pi}{8}},
\end{equation}
which, using Eq.~(\ref{EOperator}) and some trigonometric relations,
can be rewritten as
\begin{eqnarray}\label{BellOperatorSimple}
\oper{S}_{\pm} &=& \pm \sqrt{2} \left( \coper{h}_A \coper{h}_B \pm
\coper{v}_A \coper{v}_B \right) \left( \oper{h}_A \oper{h}_B \pm
\oper{v}_A \oper{v}_B \right) \nonumber \\ && \ \mp \sqrt{2} \left(
\coper{h}_A \coper{v}_B \mp \coper{v}_A \coper{h}_B \right) \left(
\oper{h}_A \oper{v}_B \mp \oper{v}_A \oper{h}_B \right) \nonumber \\
&=& \pm 2 \sqrt{2} \left[ \left( \oper{\Phi}_{\pm}^{HV} \right)^\dag
\oper{\Phi}_{\pm}^{HV} - \left( \oper{\Psi}_{\mp}^{HV}
\right)^\dag \oper{\Psi}_{\mp}^{HV} \right].
\end{eqnarray}
Since $\oper{n}_{\Phi_{\pm}} = \oper{\Phi}_{\pm}^\dag
\oper{\Phi}_{\pm}$ and $\oper{n}_{\Psi_{\pm}} =
\oper{\Psi}_{\pm}^\dag \oper{\Psi}_{\pm}$ give the probability of
observing the corresponding Bell state when evaluated in any
photon-pair state, Eq.~(\ref{BellOperatorSimple}) shows that the Bell parameter
measures the difference in occupation of pairs of Bell states,
namely $S_+ = 2 \sqrt{2} \left( n_{\Phi_+}-n_{\Psi_-}\right)$ or
$S_- = 2 \sqrt{2} \left( n_{\Psi_+}-n_{\Phi_-} \right)$.

On the other hand, the simplified Bell operator
\begin{equation} \label{simpleSOperator}
\oper{S}^\prime_{\pm} = \sqrt{2} \left( \pm \oper{\mathcal{E}}_{0,0}
+ \oper{\mathcal{E}}_{\frac{\pi}{4},\frac{\pi}{4}} \right),
\end{equation}
whose expectation value is given by Eq.~(\ref{simpleS}), can also be expressed
with the help of Eq.~(\ref{EOperator}) as
\begin{equation}\label{expandedsimpleS}
\oper{S}^\prime_{\pm} = \pm 2 \sqrt{2} \left[ \left(
\oper{\Phi}_{\pm}^{HV} \right)^\dag \oper{\Phi}_{\pm}^{HV} -
\left( \oper{\Psi}_{\mp}^{HV} \right)^\dag \oper{\Psi}_{\mp}^{HV}
\right].
\end{equation}
This establishes for the operators underlying Eqs.~(\ref{CHSH}), (\ref{simpleS}) 
the identity $\oper{S}^\prime_{\pm} = \oper{S}_{\pm}$, which implies the same relation for all
expectation values. To derive this identity, we have used the laws
of quantum mechanics:  We made use of the linear transformation
for fields or operators, Eq.~(\ref{45DegreeBasis}), that gives rise
to Malus' law, as well as the operator expectation values to connect 
Eqs.~(\ref{BellOperator}), (\ref{simpleSOperator}) to the measured quantities.
Consequently, if one accepts hidden-variable theories as ruled out by experiments
\cite{Aspect81,Aspect82,Aspect82a}, and quantum mechanics to be
valid, one can measure the Bell parameter using the simplified
expression Eq.~(\ref{simpleS}), rather than the CHSH formula Eq.~(\ref{CHSH}).

In order to elucidate the connection between 
the Bell operator $\oper{S}^\prime_{\pm} = \oper{S}_{\pm}$ and the nature of the quantum correlations,
we introduce the coincidence operator
\begin{equation}\label{PairOperator}
\oper{f}_{ab} =  \coper{a}_A \oper{a}_A \coper{b}_B \oper{b}_B.
\end{equation}
This operator measures the fraction $f_{ab} =
\aver{\oper{f}_{ab}} = \mathrm{Tr} \left( \rho \oper{f}_{ab} \right)$ of
photon pairs with polarizations $a$, $b$ observed by
Alice and Bob, respectively (e.g., $f_{hv}$ is the fraction of
events where Alice observes a horizontally polarized photon and
Bob observes a vertically polarized photon, when they both perform
measurements in the $HV$ basis). 
Using Eqs.~(\ref{BellStateOperators}), (\ref{expandedsimpleS}) and rearranging terms, 
the Bell parameter 
can be expressed in terms of the coincidences as
\begin{equation}\label{SPairs}
\frac{S^\prime_{\pm}}{\sqrt{2}} = \pm \left( f_{hh} + f_{vv} -
f_{hv} - f_{vh} \right) + \left( f_{ss} + f_{tt} - f_{st} -
f_{ts} \right).
\end{equation}
Coincidences in one basis, such as $f_{hh}$ or $f_{hv}$, can arise from either
classical or quantum correlations. 
Only quantum correlations can, however, appear simultaneously in two different bases such
as $HV$ and $ST$. 
In fact, the combination of correlations in the two bases
distinguishes completely the underlying Bell state. For instance, Eq.~(\ref{BellStatesHVST}) 
shows that the state $\ket{\Psi_{+}^{HV}}= \ket{\Phi_{-}^{ST}}$ has maximum $hv$, $vh$, $ss$, and $tt$ coincidences,
but no $hh$, $vv$, $st$, or $ts$ coincidences. Correspondingly, the quantity $S^\prime_-$ will be 
the largest in magnitude, and equal to $2\sqrt{2}$.
The relation between the Bell parameter
$S^\prime_{\pm}=S_{\pm}$ and the Bell states in the two bases, Eq.~(\ref{BellStatesHVST}), 
is summarized in Table \ref{BellTable}.

\begin{table}
\centering
\caption{Relation between Bell states and Bell parameter $S^\prime_{\pm}=S_{\pm}$.
}\label{BellTable}
\begin{tabular*}{8.6 cm}{@{\extracolsep{\fill}} c l c c c}
  \hline\hline
    &\multicolumn{1}{c}{Bell states}& $S^\prime_{+}=S_{+}$ & $S^\prime_{-}=S_{-}$&\\
  \hline
  &$\ket{\Phi_+^{HV}}=\ket{\Phi_+^{ST}}$ & $2\sqrt{2}$ & 0 & \\
  &$\ket{\Psi_+^{HV}}=\ket{\Phi_-^{ST}}$ & 0 & $2\sqrt{2}$& \\
  &$\ket{\Phi_-^{HV}}=-\ket{\Psi_+^{ST}}$ & 0 & $-2\sqrt{2}$& \\
  &$\ket{\Psi_-^{HV}}=\ket{\Psi_-^{ST}}$ & $-2\sqrt{2}$ & 0 &\\
  \hline\hline
\end{tabular*}
\end{table}

The proposed scheme proves to have smaller errors, both systematic and statistical, compared to the CHSH scheme.
We first note that the systematic error, mainly due to the inaccuracy of 
the setting of the wave plates, should be lower in most cases for the simplified scheme, 
as it involves only two combinations of angles, $\alpha^\prime_1=\beta^\prime_1=0$ and 
$\alpha^\prime_2=\beta^\prime_2=\pi/4$.
To evaluate the statistical error, we assume that the Bell
parameter is close to its maximum value $|S_\pm|=|S^\prime_\pm| \sim 2\sqrt{2}$.
We then find that, for a total of $N$ detected photon pairs, the
variance $(\Delta S)^2$ of the CHSH form of the Bell parameter $S_\pm$ 
is given by
\begin{equation}\label{varianceS}
 (\Delta S)^2=\frac{4}{N} + \mathcal{O} \left( \frac{1}{N^2} \right),
\end{equation}
while, for the simplified form $S^\prime_\pm$, we obtain the expression
\begin{equation}\label{varianceSsimple}
 (\Delta S^\prime_\pm)^2=\frac{16}{N} \left( 1- \frac{|S^\prime_\pm|}{2\sqrt{2}} \right) + \mathcal{O} \left( \frac{1}{N^2} \right).
\end{equation}
This shows that our proposed scheme provides a significantly smaller
statistical error for a Bell parameter exceeding
$\frac{3}{4}\times2\sqrt{2}$, or $|S^\prime_\pm|>2.12$.

Another possible way to measure the Bell parameter is to
observe the visibility of the interference fringes of coincidence rates vs.\ angle $\beta$ for two
fixed settings $\alpha_1=0$  and $\alpha_2=\pi/4$, and to
calculate the Bell parameter as the average visibility of the two
fringes \cite{Clauser74,Marcikic04}. 
This is only possible, however, if the state is known to be rotationally symmetric, 
and if the correlation coefficient $E(\alpha_i,\beta)$ varies sinusoidally with $\beta$. 
In contrast, our method Eq.~(\ref{simpleS}) applies generally, 
and is not restricted to the rotationally symmetric state.

In conclusion, we have pointed out an alternative scheme, within quantum mechanics, for
measuring the Bell parameter.  
The scheme requires only two different combinations of polarizer settings for Alice and Bob and 
gives smaller measurement errors compared to the standard CHSH scheme. 
We have shown how this method arises naturally from the transformation of Bell states under rotations.
This simplifies the characterization of non-classical two-photon
states, in particular for photons travelling along the same path \cite{Thompson06}. 

We would like to thank J.\ K.\ Thompson for useful discussions.
This work was supported by the NSF, DARPA, and the NSF Center for
Ultracold Atoms. J. S. acknowledges support from NDSEG and NSF.

\end{document}